\documentstyle[aaspp4]{article}

\begin{document}
\title{Far-Ultraviolet and Visible Imaging of the Nucleus of
M32\footnote{Based
on observations made with the NASA/ESA {\it Hubble Space Telescope},
obtained at the Space Telescope Science Institute, which is operated
by the Association of Universities for Research in Astronomy, Inc., 
under NASA contract NAS 5-26555.}}
\author{Andrew A. Cole,\altaffilmark{2}
John S. Gallagher, III,\altaffilmark{2}
Jeremy R. Mould,\altaffilmark{3}
John T. Clarke,\altaffilmark{4}
John T. Trauger,\altaffilmark{5}
Alan M. Watson,\altaffilmark{6}
Gilda E. Ballester,\altaffilmark{4} 
Christopher J. Burrows,\altaffilmark{7}
Stefano Casertano,\altaffilmark{8} 
David Crisp,\altaffilmark{5}
Richard E. Griffiths,\altaffilmark{8}
Carl J. Grillmair,\altaffilmark{5}
J. Jeff Hester,\altaffilmark{9}
John G. Hoessel,\altaffilmark{2}
Jon A. Holtzman,\altaffilmark{6}
Paul A. Scowen,\altaffilmark{9}
Karl R. Stapelfeldt,\altaffilmark{5}
and James R. Westphal\altaffilmark{10}}

\altaffiltext{2}{Department of Astronomy, University of Wisconsin-Madison,
475 N. Charter Street, Madison, WI 53706; cole@ninkasi.astro.wisc.edu,
jsg@tiger.astro.wisc.edu.}
\altaffiltext{3}{Mt. Stromlo and Siding Spring Observatories,
Institute of Advanced Studies, Australian National University, 
Private Bag, Weston Creek Post Office, ACT 2611, Australia;
jrm@mso.anu.edu.au.}
\altaffiltext{4}{Department of Atmospheric and Oceanic Sciences,
University of Michigan, 2455 Hayward, Ann Arbor, MI 48109.}
\altaffiltext{5}{Jet Propulsion Laboratory, 4800 Oak Grove Drive, 
Mail Stop 183-900, Pasadena, CA 91109.}
\altaffiltext{6}{Department of Astronomy, New Mexico State University,
Box 30001 Department 4500, Las Cruces, NM 88003-8001.}
\altaffiltext{7}{Space Telescope Science Institute, 3700 San Martin
Drive, Baltimore, MD 21218.}
\altaffiltext{8}{Department of Astronomy, Johns Hopkins University,
3400 N. Charles Street, Baltimore, MD 21218.}
\altaffiltext{9}{Department of Physics and Astronomy, Arizona
State University, Tyler Mall, Tempe, AZ 85287.}
\altaffiltext{10}{Division of Geological and Planetary Sciences, 
California Institute of Technology, Pasadena, CA 91125.}

\setcounter{footnote}{0}

\begin{abstract}
We have imaged the nucleus of M32 at 1600 \AA$ $ (FUV) and 5500
\AA$ $ (V) using the Wide-Field/Planetary Camera 2 (WFPC2) aboard HST.
We detected the nucleus at 1600 \AA$ $ using the redleak-free Woods
filter on WFPC2. The FUV light profile can be fit with a Gaussian of
FWHM 0\farcs46 (4.6 pixels), but cannot be resolved into individual
stars; no UV-bright nuclear structure was detected.
The (FUV$-$V) color of the nucleus
is 4.9 $\pm$0.3, consistent with earlier observations. We are 
unable to confirm any radial variation in (FUV$-$V) within 
0$\farcs$8 of the nucleus; beyond that radius the FUV surface 
brightness drops below our detection threshhold.  We also performed
surface photometry in V and found our results to be in excellent
agreement with deconvolved, WFPC1 results.  M32's 
light profile continues to rise in a nuclear cusp even within 
0\farcs1 of its center.  
No intermediate-age stellar population is required by evolutionary
population synthesis models to reproduce the (FUV$-$V) color of the 
nucleus, although these data and current models are insufficient
to resolve this issue.
\end{abstract}

\keywords{galaxies: individual (M32) --- galaxies: nuclei ---
galaxies: stellar content --- ultraviolet: stars}

\vspace{12pt}
{\it To appear in Part 1 of the Astrophysical Journal}

\newpage
 
\section{Introduction}

At a distance of just 725 kpc, M32 is the nearest ``true
elliptical'' (as contrasted to a dwarf elliptical\footnotemark)
galaxy to the Milky Way.  M32's relative proximity allowed its
outer regions to be resolved into stars by the first generation
of large optical telescopes (\cite{baa44}), while its 
crowded inner regions remained unresolved until the advent
of the Hubble Space Telescope (see \cite{gri96}).  The
ability to study both the integrated light of M32 and the
individual stars contributing to that light has made this 
galaxy a keystone of efforts to understand the stellar 
populations of elliptical galaxies beyond the Local Group.

\footnotetext{see \cite{dac97} for some illumination of
the murky waters of galaxian semantics.}

One of the outstanding puzzles in the study of elliptical
galaxies has been that of the ultraviolet excess (UVX); this
term refers to the increase in flux shortward of 2000 \AA$ $
seen in integrated spectra of elliptical galaxies, relative
to the amount predicted by the simplest model fits to their 
optical spectra (\cite{cod69}, \cite{cod79}).
An important step in identifying the stellar
population responsible for the UVX was made by \cite{bur88},
who identified a strong correlation between the strength of 
the UVX and the mean metallicity of the galaxy in a sample
of 24 quiescent systems. 

M32 defined the UV-faint, low-metallicity end of the \cite{bur88}
sample, with an M$_{1550}$ $-$ V-band color of
4.5 $\pm$0.2 and a mean metallicity of [Fe/H] $\approx$
$-0.25$ (c.f. the discussion in \cite{gri96}).  \cite{bur88}
attributed the (1550$-$V) color of M32 to the presence of
``classical'' post-asymptotic giant branch (P-AGB) stars in 
the galaxy.

The ability of P-AGB stars to produce the
(FUV$-$V) color of M32 depends strongly on the age and 
metallicity distributions of M32's stellar population.
However, the very strong degree of central concentration
which makes M32 an attractive target for studies of
integrated light has hindered progress in the characterization
of its stellar populations.  \cite{dac97} provides an
overview of the various strands of evidence for and against
the presence of an intermediate-age (few Gyr) population 
in M32. 

Spectral syntheses performed by \cite{oco80} indicated
that the dominant contributors to the present-day optical
luminosity of M32 are aged $\approx$5 Gyr with solar
metallicity, but noted that the peak star formation rate
might have occurred as long ago as 15 Gyr if the metallicity
ranged as low as one-tenth solar.  The existence of a major
intermediate-age population in M32 has yet to be proven
beyond a reasonable doubt despite the significant body of
work dealing with this problem over the past two decades.

Ground-based (e.g., \cite{fre89}, \cite{dav92}) and HST
(\cite{gri96}) photometry have both been interpreted to indicate
the presence of a significant metallicity spread in M32,
in good agreement with recent long-slit spectra (\cite{har94} 1994).
Most investigators seem to agree that significant numbers
of stars with metallicities between $-1.5 \lesssim$ [Fe/H]
$\lesssim 0.0$ are present in M32's stellar mix; of course,
the relative numbers of metal-rich and metal-poor stars are
poorly constrained due to the usual age-metallicity 
tradeoff.  \cite{har94} (1994) discovered the presence of a 
radial metallicity gradient:  between 15\arcsec$ $ and
1\arcmin$ $ from the galaxy's center, the metallicity 
decreases at the rate $\Delta$[Fe/H]/$\Delta$log(r) $=
-0.25 \pm0.07$.  An age gradient has also been reported
for M32, as \cite{gri96} infer from the integrated line
indices obtained by \cite{gon93} that the nucleus of M32
is both several gigayears younger and somewhat
more metal-rich than its outer regions.  The deepest optical
photometry of resolved stars (\cite{gri96}) suggests a median
age of 8 Gyr
for M32 with an average metallicity of [Fe/H] $= -0.25$ 
for a region 1\arcmin$ $--2\arcmin$ $ from the nucleus.

M32 was observed by \cite{ber95} in the far-ultraviolet
using HST's Faint Object Camera (FOC), but there was difficulty
in calibrating the absolute throughput of their filter set;
M32's FUV color was interpreted as an indication of a
stellar population as young as 3 Gyr with slightly sub-solar
metallicity.
\cite{bro97} also imaged M32's central region with FOC,
using the F175W and F275W filters; their results seemed
to indicate a conflict with stellar evolutionary theory,
but the interpretation of the data was complicated by
the redleak of the FOC ultraviolet filters (see, e.g., the
discussion by \cite{chi97}).  M32 was also observed by 
the Ultraviolet Imaging Telescope (\cite{ohl97});
a strong FUV$-$B color gradient was found, demonstrating that
the the galaxy's UVX is {\it weakest} in the center and
increases with radius.

We observed M32 as part of a WFPC2 GTO program to study the
hot, luminous post-AGB populations and the UV upturn phenomenon
in nearby early-type galaxies (\cite{tra94b}).
The use of the F160BW filter with WFPC2 provides a unique
constraint on the FUV properties of M32, coupling HST's 
high resolution (angular scales $\approx$ 0\farcs1 pix$^{-1}$)
with a zero-redleak (optical transmission $\ll$ 1\% 
T$_{max}$) wide-band filter that permits an
accurate determination of the (FUV$-$V) color (c.f. 
\cite{wat94}, \cite{jon96}).

\section{The Data: Observations, Reductions, and Photometry}

Images of M32 were taken on 1994 October 19--20, with the 
nucleus imaged onto the WF3 chip of WFPC2.  The data comprised
2 $\times$ 2000 sec exposures in F160BW (each exposure was split
into two parts to facilitate cosmic ray rejection), together
with a 10 sec and a 100 sec exposure in F555W. 
F160BW has a response from roughly 
1200--2100 \AA$ $ and is characterized by an extremely low throughput,
while F555W is the WFPC2 analog to Johnson V.
WFPC2 and its filter set are described in detail in \cite{tra94} and
\cite{bir96}.  The images were reduced in the standard way according
to the procedures of \cite{hol95a}; cosmic ray rejection and 
image manipulations were performed within IRAF\footnotemark.

\footnotetext{IRAF is distributed by the National Optical 
Astronomical Observatories, which are operated by the Association
of Universities for Research in Astronomy, Inc., under cooperative
agreement with the National Science Foundation.}

We identified bright stars in the outer regions of the nearby
spiral M31 in the V and FUV frames and used their relative offsets
to check the alignment of our WFPC2 images, which proved to be aligned
at the sub-pixel level.  The central 15\arcsec$ $ $\times$ 15\arcsec$ $
of M32, as seen at 1600 \AA, are shown in Fig. 1; the galaxy is
lurking in the shadows of detectability.  A radial ultraviolet
light profile, derived with the IRAF task {\tt IMEXAM}, is shown 
in the upper left.  The broad, shallow profile differs strongly 
from the high-amplitude peaks produced by the unremovable residue
of large cosmic ray events such as the one several arcseconds 
East of the galactic nucleus.  In contrast,
application of {\tt IMEXAM} to random locations in the field
typically produced constant value or linear fits with no
central peak.

Photometry was performed according
to the procedure laid out in \cite{hol95b}, using an 0\farcs5
aperture and their tabulated zeropoints for the STMAG system.
Zeropoint corrections were made to account for the deposition
of solid contaminants on the cold window of the CCD and for 
chip-to-chip variations in sensitivity, following \cite{hol95b}.
\cite{bur88} report
an interstellar reddening of A$_B$ = 0.31 mag towards M32; to
correct for this, we assumed a standard Galactic extinction law
with R$_V$ = 3.1 (\cite{ccm89}).  A$_{555}$ was then taken from
\cite{hol95b} to be 0.25 mag (assuming a K5 stellar spectrum).
The selective extinction in 
F160BW was calculated using the method of \cite{col97} to 
be A$_{160}$ = 0.70 mag for M32.

\section{The UV Color of the Nucleus}

An automated search for point sources near the nucleus of M32
proved negative; visual inspection of the nuclear region 
suggested the presence of a faint, diffuse contribution to 
the FUV light.  This impression was strengthened when we found
that the best Gaussian fit to the diffuse light peaked at the
position of the optical center of the galaxy, with a FWHM = 4.6
pix (compare to the F160BW point-spread function with FWHM 
$\approx$ 2.1 pix). For display purposes, we convolved the
F160BW image with a 5$\times$5 median filter in order to
sharpen the contrast between diffuse features and the background.
The filtered image is shown in Fig. 2; we have plotted
isophotes of the F555W image on top of the F160BW image in order
to demonstrate that the emission we have identified with the nucleus
is indeed coincident with the optical center of the galaxy.

Aperture photometry of the nucleus yielded $m_{160}$ = 18.39 
$\pm$0.26 mag (1$\sigma$ random photometric error; systematic
effects are expected to create an additional $\approx$10\%
uncertainty).  To quantify the significance of this measurement,
we placed 120 identical 0\farcs5 apertures at random across the
combined WF3 frame and examined the results.  Sky values were
determined locally in all cases, and the mean value was not
found to vary appreciably across the field.  The resulting
photometry showed that M32's nucleus is the brightest region
of the WF3 frame. A comparison to the distribution
of test aperture magnitudes showed that the nucleus sits
3.9$\sigma$ above the background, defined by the mean number
of counts in the test apertures and the scatter around that mean.
This number is likely an {\it underestimate} of
the significance of the detection, because the mean of the 
test aperture counts was not corrected for contamination
by unremoved cosmic ray events.  We determine from the test
aperture count distribution that our minimum 3$\sigma$ detection
level occurs at $m_{160}$ = 18.7.

In contrast, the nucleus is extremely obvious in the V-band;
the diffuse optical light has an azimuthally averaged FWHM of
4.2 pix, and a total magnitude $m_{555}$ = 12.99 $\pm$ 0.01.
Thus we find a dereddeded (160$-$555) color of 4.9 $\pm$ 0.3
for the nucleus of M32.  This is consistent with the preliminary
analysis of the same dataset by \cite{jon96} which yielded 
(160$-$555) = 4.7 $\pm$ 0.2, and in satisfactory agreement with the 
IUE (1550$-$V) color of 4.50 reported by \cite{bur88}.  Varying
the aperture radius between 0\farcs1 and 0\farcs8 uncovered 
no variation in (160$-$555), although the errors are large
enough to hide a mild gradient if it exists.  Beyond 0\farcs8,
the FUV surface brightness drops below our detection threshhold.

\section{Surface Photometry of the Nuclear Region}

We performed surface photometry within 5\arcsec$ $
of the nucleus  in the F555W frames.  Our results agree with the
WFPC1 results of \cite{lau92}.  We find a surface
brightness at 0\farcs3 of $\Sigma_{555}$ = 12.5 mag arcsec$^{-2}$,
rising to 11.9 mag arcsec$^{-2}$ at 0\farcs1, in excellent
agreement with both \cite{lau92} and the model mass distribution
derived by \cite{van97a}.  The F555W major axis radial profile of the
central 5\arcsec$ $ of M32 is shown in Figure 3.  M32 shows a 
pronounced nuclear cusp in its light profile.

We did not obtain sufficient signal-to-noise in F160BW to
perform accurate surface photometry, but the FUV light of
M32's nucleus is clearly more broadly distributed than a 
point source.  We infer from our agreement with the IUE
(1550$-$V) color the the FUV light is distributed similarly
to the optical light.  The smoothly distributed light
indicates that it is likely due to the presence of a 
relatively large number of faint objects.

\section{Discussion}

It seems impossible to reproduce the integrated 
spectrum of M32 with a single-age, single-metallicity
``simple stellar population'' (SSP) (\cite{har94} 1994).
However, SSPs remain a useful tool for interpreting the
(160$-$555) color, because they provide the tools
with which to construct more realistic models.  For
SSPs, the (FUV$-$V) color is predicted to be a 
sensitive function of both age and metallicity, and
hence a valuable observation for the derivation of
star formation histories (\cite{bre94}).  The
exact dependencies of (FUV$-$V) on stellar age and metallicity
are highly uncertain, and this limits its predictive power. As
discussed in detail by \cite{yi97}, the age predictions depend
critically on poorly modeled processes such as stellar mass-loss.
Therefore, the uncertainties in our interpretation are dominated
by systematic model effects rather than observational errors.

In order to provide an interpretational starting point,
we compared the observed (160$-$555) color of the 
nucleus of M32 to the SSP predictions of \cite{tan96}.
Table 1 gives the ages of SSPs which are predicted
to exhibit (160$-$555) = 4.9 $\pm$0.3, for a range 
of metallicities.  In Table 1, $t_y$ indicates the
acceptable age range for models in which the FUV light
derives from main-sequence stars, while $t_o$ gives
the ages for which evolved stars provide the FUV light.
The errorbars in Table 1 are {\it internal, random, observational}
errors only, and merely reflect the propagation of photometric
error into the model predictions. A thorough discussion of
the systematic errors in these models is beyond the scope of
this paper (c.f., \cite{yi98}), but uncertainties in the 
physics of stellar-mass loss and helium-enrichment limit the
absolute precision of such models to several billion years
(particularly the older models, which most concern us here).
The ages in Table 1 are therefore most useful as relative
measures, so that M32's stellar populations may be compared
with those of other galaxies using a consistent set of models.
As absolute measures, they are subject to extreme uncertainty.

In these models, the ratio of FUV to optical light
is not a monotonically decreasing function of age: early
in a galaxy's lifetime, (FUV$-$V) increases (gets redder)
as the hot upper main-sequence stars die off.  Eventually,
as the total luminosity of the SSP continues to fade with
age, the FUV contribution of the short-lived P-AGB stars
begins to become important, and (FUV$-$V) becomes bluer
again.  At sufficiently high metallicities (Z $>$ 2Z$_{\sun}$),
these models predict a point at which low-mass helium-burning stars
avoid the AGB in favor of long-lived UV-bright phases,
causing a sharp increase in the FUV flux as this behavior 
sets in at ages $\approx$5--10 Gyr.

The (160$-$555) color of M32 can be matched by either 
young or old SSPs.  We rule out the young (0.5--2 Gyr)
class of models for two reasons: first, these ages are
much younger than even the youngest intermediate-age
populations inferred from optical spectroscopy and 
photometry of M32; and second, models of P-AGB stars
(\cite{vas94}) for such young ages are predicted to
have $m_{160}$ $\approx$ 18.5 at the distance of M32;
only one such P-AGB star would be required to account
for the FUV light of M32's nucleus, and would be seen
as a point source in our images.  The remaining models indicate an
SSP age of 12--13 Gyr for M32's nucleus for metallicities
less than about 5 times solar.  In this type of SSP,
the FUV light is produced by classical P-AGB stars.  
In the super metal-rich case the age of the SSP drops
to $\approx$6 Gyr due to the production of hot
AGB-manqu\'{e} stars via metal-enhanced mass loss on 
the RGB (\cite{tan96}, \cite{dor95}, \cite{gre90}).
It is important to remember that these age values are subject
to large, model-driven systematic errors and should not be
considered as definitive determinations.

M32 is not a likely candidate to harbor AGB-manqu\'{e}
stars, as the required metallicity is too high by 
a factor of two or more.  The best matching SSPs of
reasonable metallicities are then 12--13 Gyr old; this
is consistent with the results of \cite{gri96} indicating
that half the optical light of M32 comes from stars older
than $\approx$8 Gyr.  Note that M32 is {\it not} an SSP;
any intermediate-age population that contribues to the
optical light is predicted to be FUV-faint and thus
pushes the required age of the FUV-bright population
to even older ages in order to reproduce the integrated
(160$-$555) color.

We do not have enough radial coverage in F160BW to
permit comparison with the UIT result that M32's
FUV light falls off more slowly with radius than
its optical light (\cite{ohl97}).  No gradient in
(160$-$555) is seen with the central 0\farcs8 of
M32's nucleus.  If the UIT result is interpreted
in the context of SSPs, it would suggest that
either M32's metallicity is increasing with radius
(contrary to previous results of \cite{har94} 1994),
or that the galaxy's mean age is increasing with
radius.  This type of radial age gradient has also been
suggested to exist in the dwarf elliptical galaxy
NGC 147 (\cite{han97}) and in the Andromeda I dwarf
spheroidal (\cite{dac96}); it may turn out to be a
general feature of small, spheroidal galaxies.

\section{Summary}

$\bullet$ We have imaged the nucleus of M32, an elliptical galaxy with 
a weak ultraviolet upturn, in the visible and far-ultraviolet
with WFPC2 aboard the {\it Hubble Space Telescope}.  These data extend
FUV observations of M32's nucleus to smaller radii than have heretofore
been possible.  We found the nucleus to have a color (160$-$555) = 
4.9 $\pm$ 0.3, in agreement with the results from IUE.

$\bullet$ Like \cite{ber95}, we were unable to resolve the FUV
light of M32 into individual stars.  Unlike \cite{ber95} and
\cite{bro97}, we do not require the presence of a 3--5 Gyr population
or AGB-manqu\'{e} stars to explain the (FUV$-$V) color of M32.
Using the simple stellar population models of \cite{tan96}, we
find that the (160$-$555) color of M32 is most easily explained
by the presence of a moderately metal-poor stellar population 
aged $\gtrsim$12 Gyr.  Given the large uncertainties in the
processes which can induce old, low-mass stars to high-temperatures,
this age should be considered uncertain to within $\pm$ $\sim$5 Gyr
(Yi {\it et al.} 1997, 1998); thus an intermediate-age population
cannot be ruled out.

$\bullet$ Surface photometry in the visible is in excellent
agreement with deconvolved WFPC1 and ground-based data
(e.g., \cite{lau92}).  Because of the poor signal-to-noise ratio,
there is no evidence in our data that the FUV light is
distributed differently than the optical light.

\acknowledgments

This research was carried out by the WFPC2 Investigation Definition Team
for JPL and was sponsored by NASA through contract NAS 7--1260.

{}

\newpage

\begin{center}
{\bf Table 1.} Age ranges which reproduce the observed
(160$-$555) color of the nucleus of M32, for five different
metallcities.  The colors are derived from the models
of Tantalo {\it et al.} 1996.  In the `young' SSPs, the
UV light is provided by the main-sequence, while the
'old' SSPs derive their UV colors from evolved stars.
Errorbars derive from propagation of our photometric
errors into the models; additional systematic errors
of order $\pm$5 Gyr (Yi {\it et al.} 1998) should be applied to the
old SSP solutions (Column 3).
\vspace{0.5in}

\begin{tabular}{|c|c|c|} \hline
Z & $t_y$ (Gyr) & $t_o$ (Gyr) \\ \hline
0.004   & 2.6 $\pm$0.4  & 13.1 $\pm$0.4  \\
0.008   & 1.7 $\pm$0.2  & 12.3 $\pm$0.3  \\
0.02    & 1.1 $\pm$0.1  & 13.0 $\pm$0.4  \\
0.05    & 0.69 $\pm$0.04 & 12.6 $\pm$0.6 \\
0.10    & 0.62 $\pm$0.05 & 5.9 $\pm$0.1  \\ \hline
\end{tabular}
\end{center}

\newpage

\begin{figure}
\plotone{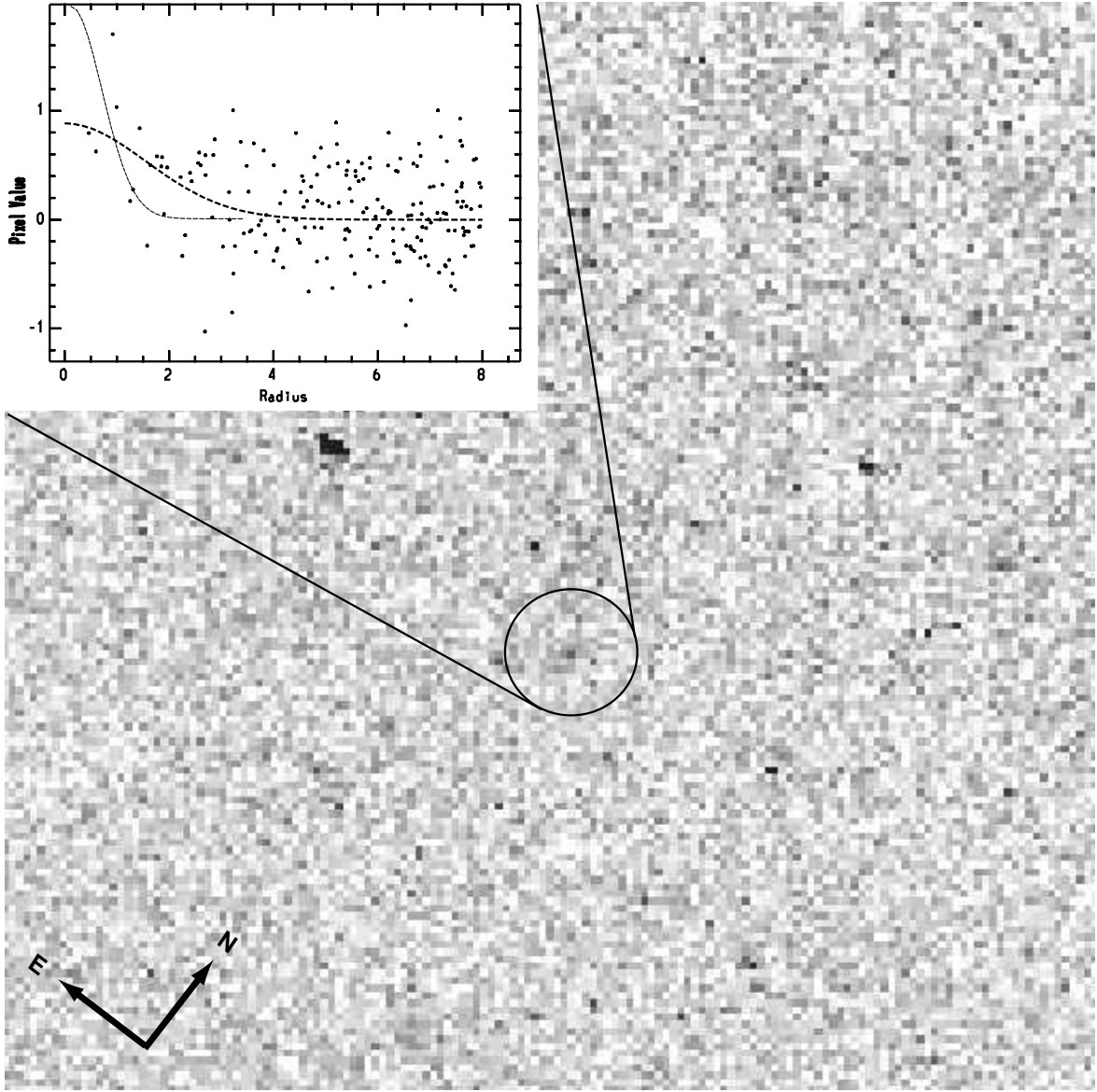}
\caption{15\arcsec$ $ $\times$ 15\arcsec$ $ of the
2$\times$2000 sec F160BW image of M32, centered on the galaxy's 
nucleus as determined from the F555W images.  North and East are
indicated on the figure.  The galaxy is not immediately obvious
in the far-ultraviolet image, but proves to be detectable at the
3.9$\sigma$ level.  At upper left, we have plotted a fit to the 
radial profile of the nucleus (heavy dashed line); for comparison, a
point-source profile has been scaled to the same total flux and
overplotted on the data (narrow dashed line).  Sharply peaked
cosmic-ray residuals can be seen across the image (e.g., the
dark region 5\arcsec$ $ E of the nucleus).}
\end{figure}

\begin{figure}
\plotone{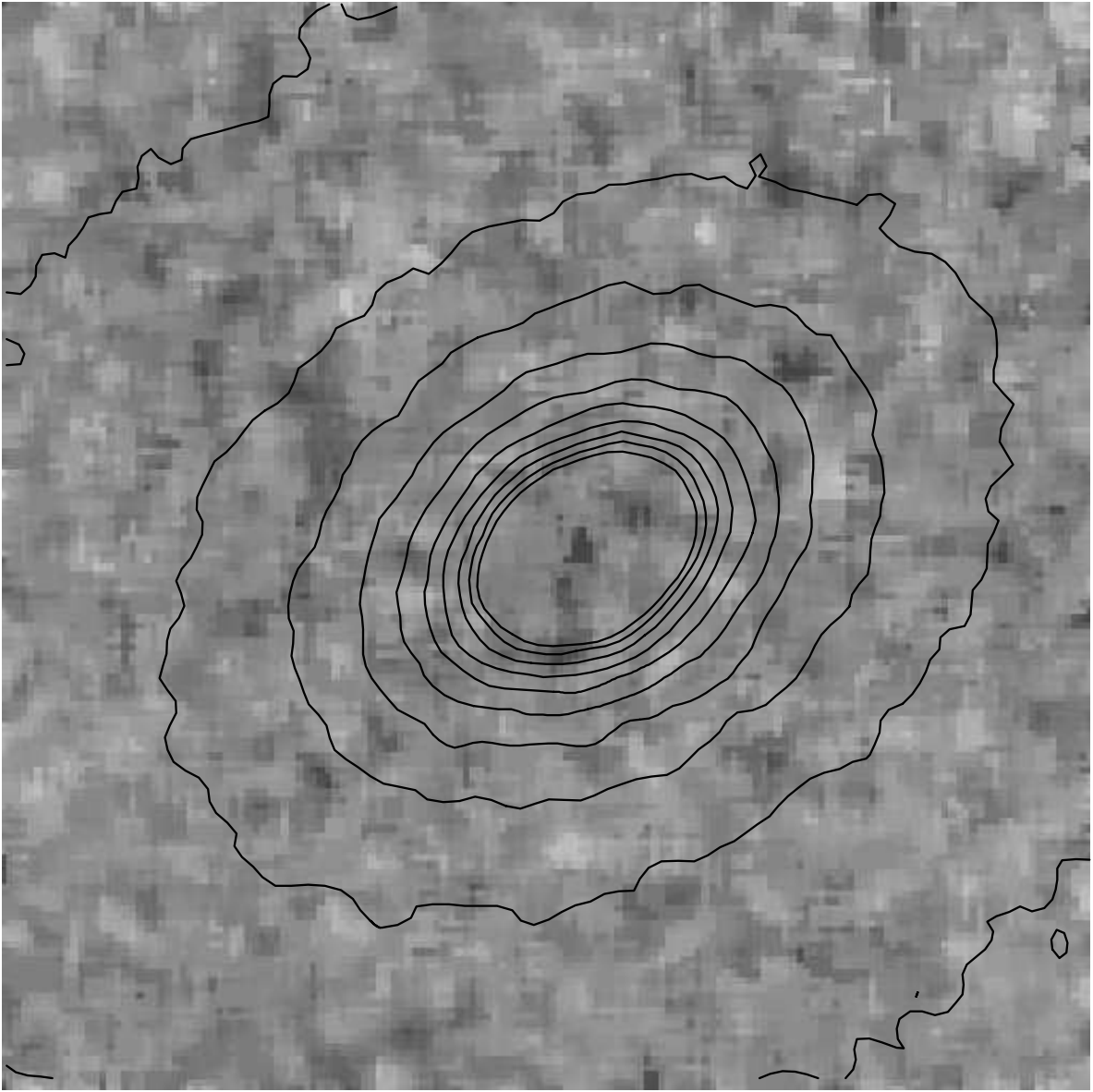}
\caption{The F160BW image from Fig. 1, after
median filtering to enhance the visibility of extended sources.
In the median-filtered image, the nucleus has risen up out of the 
noise, and is seen to correspond exactly to the center of the 
galaxy.  F555W isophotal contours are overplotted in black.}
\end{figure}

\begin{figure}
\plotone{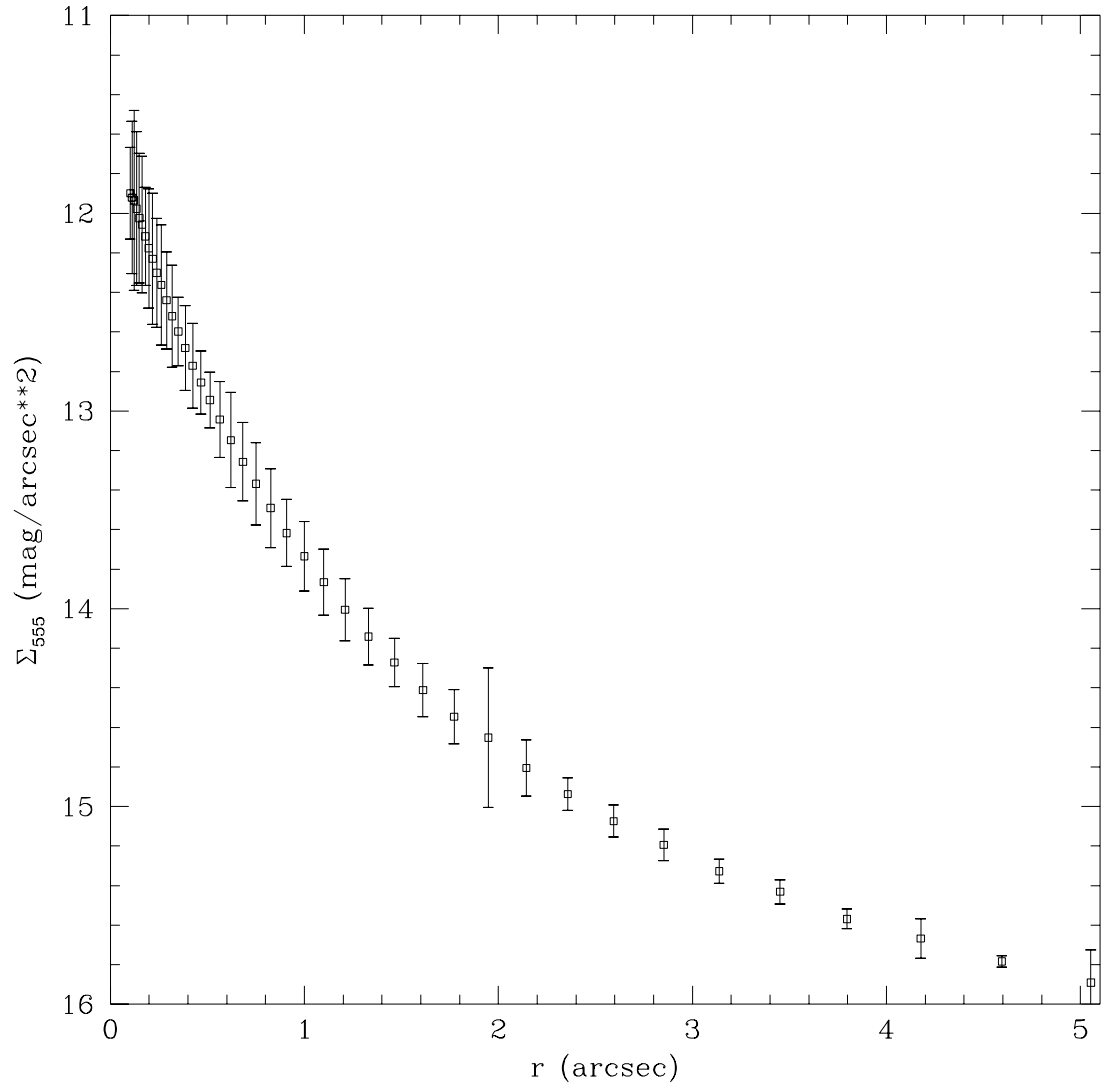}
\caption{Major-axis radial profile of M32 as
seen in the F555W filter.  No deconvolution has been applied.
The results are in excellent agreement
with earlier determinations, and extend to within 0\farcs1 of 
M32's center.}
\end{figure}

\end{document}